\documentclass[conference]{IEEEtran}
\IEEEoverridecommandlockouts

\usepackage{cite}
\usepackage{comment}
\usepackage{amsmath,amssymb,amsfonts}
\usepackage{graphicx}
\usepackage{textcomp}
\usepackage{xcolor}
\usepackage{mathtools}
\usepackage{algpseudocode}
\usepackage{algorithm}
\usepackage{algorithmicx}
\usepackage{textcomp}
\usepackage{filecontents}
\usepackage{microtype}
\usepackage{multirow}
\usepackage{multicol}
\usepackage{float}
\usepackage{adjustbox}
\usepackage{booktabs, make cell, tabularx}
\usepackage{url}
\usepackage{booktabs}
\usepackage{subfigure}

\newcolumntype{C}[1]{>{\centering\arraybackslash}p{#1}}
\newcolumntype{L}{>{\raggedright\arraybackslash}X}
\usepackage{siunitx}
\usepackage[utf8]{inputenc}
\renewcommand{\thetable}{\arabic{table}}
\usepackage{etoolbox}
\usepackage{xparse}
\makeatletter

\def\BibTeX{{\rm B\kern-.05em{\sc i\kern-.025em b}\kern-.08em T\kern-.1667em\lower.7ex\hbox{E}\kern-.125emX}}
\begin{document}

\title{Python-based DSL for generating Verilog model of Synchronous Digital Circuits
{\footnotesize \textsuperscript{}}}


\author{
 \IEEEauthorblockN{Mandar Datar, Dhruva S. Hegde, Vendra Durga Prasad, Manish Prajapati, Neralla Manikanta,\\ Devansh Gupta, Janampalli Pavanija, Pratyush Pare, Akash, Shivam Gupta, and Sachin B. Patkar}
 \IEEEauthorblockA{Department of Electrical Engineering\\ Indian Institute of Technology Bombay, India\\
 Email:\{mandardatar, patkar\}@ee.iitb.ac.in}
}

\maketitle

\begin{abstract}
We have designed a Python-based Domain Specific Language (DSL) for modeling synchronous digital circuits. In this DSL, hardware is modeled as a collection of transactions -- running in series, parallel, and loops. When the model is executed by a Python interpreter, synthesizable and behavioural Verilog is generated as output, which can be integrated with other RTL designs or directly used for FPGA and ASIC flows. In this paper, we describe - 1) the language (DSL), which allows users to express computation in series/parallel/loop constructs, with explicit cycle boundaries, 2) the internals of a simple Python implementation to produce synthesizable Verilog, and 3) several design examples and case studies for applications in post-quantum cryptography, stereo-vision, digital signal processing  and optimization techniques. In the end, we list ideas to extend this framework.
\end{abstract}


\section{Introduction}
Current high-level synthesis (HLS) tools such as Xilinx Vivado HLS, Intel HLS, etc. convert `high-level un-timed C/C++ model' into synthesizable Verilog code \cite{cardoso}. A tool like Bluespec-SystemVerilog \cite{bsvcite} expects the user to break the computation into `atomic actions' and the compiler emits synthesizable Verilog after scheduling the actions correctly and optimally. If a project is entirely developed using such tools, one can debug (edit-compile-debug) at high-level C/C++/BSV itself. However, if one has to interface the design with existing RTL modules (e.g. third party), RTL/Verilog simulation needs to be carried out. Signal-level debugging is not easy when the RTL itself is machine-generated since most of the signal names in the RTL can not be correlated with variable names in the high-level model. With HLS tools, the user can influence the RTL generation using pragmas (e.g. to unroll/pipeline loops) however, final scheduling is done by the tool.  There are also tools like PyMTL \cite{pymtl}, and MyHDL \cite{myhdl}, which convert a logic design expressed in a high-level language (Python) to RTL. The DSL described in this chapter is not an HLS tool. It is at a lower level than HLS, but it is at a higher level compared to RTL. It is extremely light-weight, easy to use and generates synthesizable Verilog with good readability.\\ 
In the Python-based DSL presented in this paper:
\begin{itemize}
  \item Actual computation (comparisons, addition, multiplication, etc.) is written as Python expressions.
  \item Python statements are grouped into `leaf sections', where the entire computation under one leaf happens in a single clock cycle.
  \item Series, parallel and loop sections group other such sections forming a tree.
  \item Generated Verilog is behavioural and preserves register/wire names and expression structure from the user's Python code.
  \item Full Python facilities are available for static elaboration.
\end{itemize}

\begin{figure}[h]
  \centering
  \includegraphics[width=3.3in]{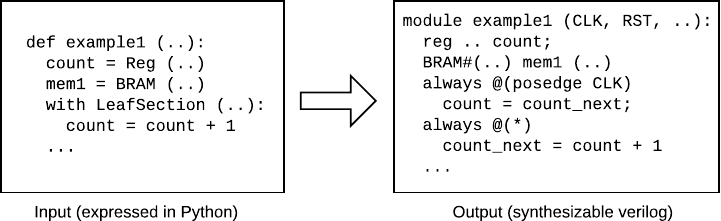}
  \caption{Python to Verilog}
  \label{hlbd}
\end{figure}

\noindent
Figure \ref{hlbd} provides a high-level picture of the Python DSL.

The flow of this paper is as follows. Section 2 introduces the proposed DSL framework and its constructs. Section 3 elaborates on how Python language features are used to build the DSL. This is followed by Section 4, which presents several complex examples of hardware designed using the DSL and the results obtained. Section 5 outlines the future directions of this work. Finally, Section 6 concludes the paper.

\section{Python DSL Constructs}

In the Python DSL model, the computation is broken into chunks called LeafSections. The entire body of one LeafSection `executes' in one clock cycle.
For each module, clock, reset, enable-ready for inputs and outputs signals are implicitly present in the generated Verilog.\\

\noindent
The statements that are written under SerialSections are executed in sequence. This construct can be used for sequential modelling.
\begin{footnotesize}
\begin{verbatim}
with SerialSections ("S1"):
  with LeafSection ("a"):
    display ("running 'a'")
  with LeafSection ("b"):
    display ("running 'b' after 'a'")
  with LeafSection ("c"):
    display ("running 'c' after 'b'")
\end{verbatim}
\end{footnotesize}

\noindent
Similarly, the statements that are written under ParallelSections are executed in parallel. This construct can be used for concurrent modelling.
\begin{footnotesize}
\begin{verbatim}
with ParallelSections ("P1"):
  with LeafSection ("a"):
    display ("running 'a'")
  with LeafSection ("b"):
    display ("running 'b' with 'a'")
\end{verbatim}
\end{footnotesize}

\noindent
The statements under WhileLoopSections and ForLoopSections are executed iteratively. 
\begin{footnotesize}
\begin{verbatim}
with ForLoopSection ("F1", "i", 0, 3):
  with LeafSection ("a"):
    display ("i=%0d, running 'a'", i)
  with LeafSection ("b"):
    display ("i=%0d, running 'b'", i)
\end{verbatim}
\end{footnotesize}

\noindent
All the constructs described can be combined (i.e. LeafSections can be arranged to run serially, in parallel, in a nested manner, and in loops) to model complex algorithms.\\

\noindent
The Python DSL has mainly two types of program variables - {\tt Reg} and {\tt Var}. They both can be of arbitrary bit-width.
\begin{itemize}
    \item {\tt Reg} variables are used to synthesize registers. The values assigned to variables of type {\tt Reg} are updated in the next clock cycle.

    \item {\tt Var} variables are used to synthesize pure combinational logic. The values assigned to variables of type {\tt Var} are updated in the same clock cycle.
\end{itemize}
{\tt RegArray} is also supported, which synthesizes an array of registers interfaced with decoders at the inputs and multiplexers at the outputs.

\section{Construction of the DSL}

This section explains the construction of the Python DSL.

\subsection{Symbolic expressions and context blocks}

This section shows the Python language features used in the proposed DSL framework. These features keep the statement syntax simple and build the section tree when the Python model is executed. Operator overloading is used to build symbolic expressions from user-written Python expressions \cite{sympy, pyeda}.\\

A `{\tt symbol}' base class is created to represent register, port and wire objects. As shown below, overloaded operators build `{\tt BinaryExpression}' objects.

\begin{footnotesize}
\begin{verbatim}
 1 class BinaryExpression:
 2     def __init__ (self, op, a, b):
 3       ...
 4     def to_string (self):
 5       return '(' + self.a.to_string () 
 6                  + self.op 
 7                  + self.b.to_string () + ')')
 8 
 9 class Symbol:
10     def __init__ (self, name, width):
11       ...
12     def to_string (self):
13       return self.name
14     def __add__ (self, other):
15       return BinaryExpression ('+', self, other)
16     ...
\end{verbatim}
\end{footnotesize}

On line 14, the {\tt +} operator is overloaded. It prepares a {\tt BinaryExpression} object, referring to the operands {\tt self} and {\tt other}. The following code indicates how a symbolic expression is built when Python code is executed.

\begin{footnotesize}
\begin{verbatim}
a = Symbol ("a", 32)
b = Symbol ("b", 32)
c = Symbol ("c", 32)

result = a + b + c
\end{verbatim}
\end{footnotesize}

\noindent
Here, `result' is a label pointing to a symbolic expression object.

\begin{footnotesize}
\begin{verbatim}
>>> print (result.to_string ())
((a + b) + c)
\end{verbatim}
\end{footnotesize}

\noindent
Python language has a construct called `` `{\tt with}' expression'' which is used for opening a file, processing it with a block of statements,
and closing the file automatically. This construct is used to demarcate the boundaries of the Sections.

\begin{footnotesize}
\begin{verbatim}
class ContextClass:
  def __enter__ (self):
    print ("Entering.")
  def __exit__ (self, t, value, traceback):
    print ("Exiting.")

c = ContextClass ()

print ("statement 1")
with c:
    print ("statement 2")
    print ("statement 3")
print ("statement 4")
\end{verbatim}
\end{footnotesize}

\noindent
The above block of code produces:

\begin{footnotesize}
\begin{verbatim}
statement 1
Entering.
statement 2
statement 3
Exiting.
statement 4
\end{verbatim}
\end{footnotesize}

\noindent
Inside the {\tt \_\_enter\_\_} method, we add a new Section object as root, or add it to an already open section object. As the body of the section executes, we add child sections/statements to this new section object. In the {\tt \_\_exit\_\_} method, we close the section object.\\

\subsection{Building a tree of section objects}

With this background (symbolic expressions and context objects), we proceed to explain the construction of module objects, which will contain
a tree of sections (series/parallel/loop etc.), and each LeafSection will hold a block of statements.

\begin{footnotesize}
\begin{verbatim}
 1  class example1(HWModule):
 2    def __init__ (self, instancename, ..):
 3      HWModule.__init__ (instancename)    # [
 4      c = Reg ("c", 32)
 5      with LeafSection ("L1"):
 6        Assignment (c, c + 1)
 7      self.endModule ()                   # ]
 8
 9  e = example1 ("e", ..)
10  e.emitVerilog ()
\end{verbatim}
\end{footnotesize}

\noindent
On creation of object `e' at line 9, \_\_init\_\_ method gets called automatically. Inside example1.\_\_init\_\_, its base class is called \_\_init\_\_ method (line 3). It will record the start of the definition of a new hardware module. The statement on line 4 will create a symbol object and its constructor will record it as a member of the current hardware module. On entering the LeafSection `L1', a section object gets added to the tree of sections of the current hardware module. When the statements within the LeafSection `L1' are executed, e.g. the assignment statement on line 6, it will record the LHS and RHS expressions (which will be of type Symbol/BinaryExpression etc.) into the code block associated with currently open LeafSection object. On exiting from the LeafSection `L1', the object in the tree will be closed, so that, any subsequent `with LeafSection..' will add a new child to the tree of sections. On line 7, endModule () statement will mark the end of the definition of 
the current hardware module.\\
Thus, the object e will contain a Register member `c', and a tree of sections, containing one LeafSection `L1', with a single assignment statement in it. On line 10, the base class i.e. HWModule.emitVerilog method is called. It will emit Verilog code defining the module example1.\\

\subsection{Conversion of Python to Verilog}
A simple example having two LeafSections in series is provided for illustration, with the intended state diagram, and parts of the generated Verilog.

\begin{footnotesize}
\begin{verbatim}
@hardware
def add_sub (a, b, c):
    a = RegIn  ("a", 32)
    b = RegIn  ("b", 32)
    c = RegIn  ("c", 32)
    d = RegOut ("d", 32)
    tmp = Reg  ("tmp", 32)
    with LeafSection ("add"):
      tmp = a + b
      display ("add: a=%0d, b=%0d", a, b)
    with LeafSection ("sub"):
      d = tmp - c
      display ("result: %0d", tmp - c)
\end{verbatim}
\end{footnotesize}

\noindent
Here, `@hardware' is a transformer, that replaces the definition of `add\_sub' with a class similar to `example1' above. In particular, the original contents of `add\_sub' become part of the `\_\_init\_\_' method of the class.\\

\noindent
In Python, AST (abstract-syntax-tree) can be easily modified. The AST for the code inside `with Section*' blocks are re-written and the assignment statements are updated.

\begin{footnotesize}
\begin{verbatim}
    tmp = a + b
\end{verbatim}
\end{footnotesize}

\noindent
The above code is translated to

\begin{footnotesize}
\begin{verbatim}
    Assignment (tmp, a + b)
\end{verbatim}    
\end{footnotesize}

\begin{figure}[h]
  \centering
  \includegraphics[width=2.75in]{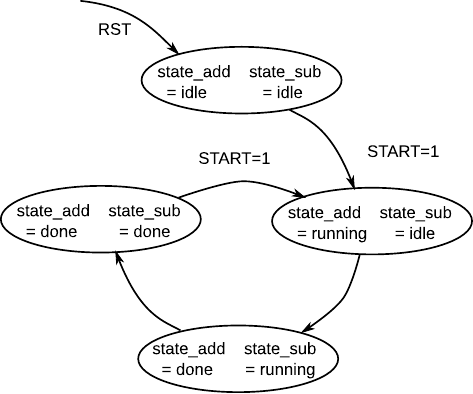}
  \caption{A state-diagram for `LeafSections' in `add\_sub' module}
\end{figure}

\noindent
Parts of the behavioural Verilog code emitted by the framework are listed below.

\begin{footnotesize}
\begin{verbatim}
module add_sub (
  CLK, RST,
  START, Done, get_done, Ready,
  a, b, c, d
  );
  ...
  reg [1:0] state_add = 2'd0;
  reg [1:0] state_add_WIRE;
  reg [1:0] state_sub = 2'd0;
  reg [1:0] state_sub_WIRE;
  ...
  always @(*) begin
  ...
    if (state_add == 1) begin
      tmp_WIRE = (a_inreg + b_inreg);
      state_add_WIRE = 2;
      state_sub_WIRE = 1;
    end
    if (state_sub == 1) begin
      d_outreg_WIRE = (tmp - c_inreg);
      state_sub_WIRE = 2;
      state_st_WIRE = 2; 
      ...
    end
  ...
  end
  ...
endmodule
\end{verbatim}
\end{footnotesize}

A testbench can be written in Python itself as:

\begin{footnotesize}
\begin{verbatim}
 1 @hardware
 2 def my_tb():
 3  m1 = add_sub ("m1", None, None, None)
 4  with SerialSections ("S"):
 5    with LeafSection ("S10"):
 6      m1.start (a=Const (32,21), b=Const (32,34), 
 7                c=Const (32,5))
 8    with LeafSection ("S11"):
 9      display ("Result = [%0d]", m1.isDone ()[0])
\end{verbatim}
\end{footnotesize}

\noindent
Here, on line 3, the `add\_sub' module is instantiated. In this way, \textit{hierarchy of modules} can be created. Note that, the hierarchy is not flattened. Each module is emitted separately i.e. the hierarchy is preserved in the generated Verilog. So, in this example, {\tt my\_tb} module will declare an instance of {\tt add\_sub} as {\tt m1}.
And the definition of {\tt add\_sub} will be emitted as a separate module.

\subsection{Simulation and Synthesis} 

\begin{figure}[!ht]
  \centering
  \includegraphics[width=3in]{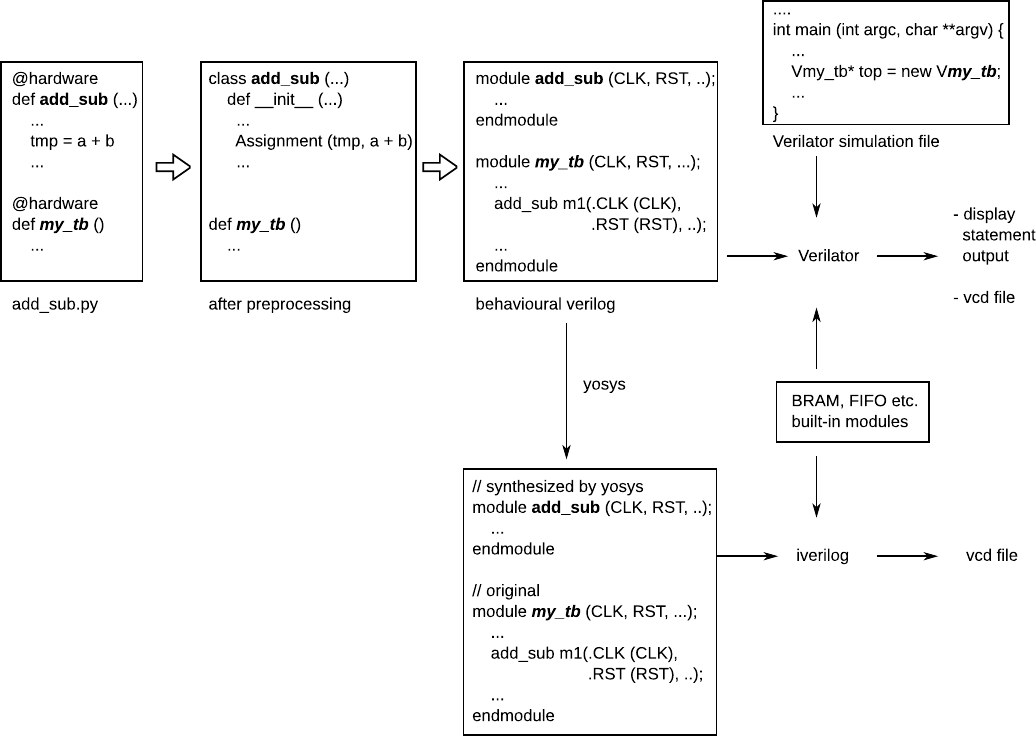}
  \caption[Python DSL : Overall Flow]{Python DSL : Overall Flow}
  \label{fig:hlsflow}
\end{figure}

As shown in Figure \ref{fig:hlsflow}, user-written Python code is passed through preprocessing and Verilog generation stages. Generated Verilog is simulated using an RTL simulator (e.g. Verilator).
A few more Verilog files such as BRAM and FIFO are also provided to the simulator. The simulator produces a vcd file and prints results of display statements specified by the user in Python. Generated Verilog is then given to Yosys tool for synthesis. The BRAM module is written in Verilog, and Yosys infers FPGA BRAM instance when `memory -bram' pass is run.

\noindent
Even though the basic modules are available as Verilog modules, while building the module hierarchy in Python, a Python model/interface is needed to represent them. The Python interface model for the basic FIFO module looks like this:

\begin{footnotesize}
\begin{verbatim}
 1 class Fifo (HWModule):
 2   def __init__ (self, instancename, width):
 3     ...
 4     self.addParameter ("width", width)
 5     ...
 6     self.addOutPort ("read_data", width)
 7     self.addInPort  ("read_enable",  1)
 8     ...
 9   def read (self):
10     addCondition (self.getPortWire ("read_ready")
                                     == Const (1,1))
11     Assignment (self.getPortWire ("read_enable"), 
                                        Const (1,1))
12     return self.getPortWire ("read_data")
13   def write (self, data):
14       ...
\end{verbatim}
\end{footnotesize}

\noindent
On line 4, a parameter `width' is defined. The symbolic variable for `read\_data' wire will get that width. On line 10, a condition "read\_ready == 1'b1" is added. Whenever a `read' function on a FIFO instance is called inside a leaf section, the body of the leaf section will be wrapped
inside this `if condition' i.e. that LeafSection will `execute', and trigger the execution 
of subsequent LeafSections only when this condition is satisfied.\\

\noindent
Python facilities can be used for static elaboration.
\begin{footnotesize}
\begin{verbatim}
 1     a = Reg ("a", 32) 
 2     with LeafSection ("loop1"):
 3       tmp = Var ("tmp", 5)
 4       for i in range (32):
 5         tmp = tmp + a[i]
\end{verbatim}
\end{footnotesize}

\noindent
The pre-processing stage preserves the for-loop on line 4. Note that this is Python's built-in `for' loop, and not `with ForLoopSection' defined by the DSL framework. When this code is run for Verilog generation, the loop gets expanded by Python, making it equivalent to:

\begin{footnotesize}
\begin{verbatim}
         tmp = tmp + a[0]
         tmp = tmp + a[1]
         ...
         tmp = tmp + a[31]
\end{verbatim}
\end{footnotesize}

\noindent
Similarly, all facilities available in Python are available to the user for static elaboration.

\subsection{Interface Generation}

Apart from the standard FIFO interface, the Python DSL may be ported to generate Verilog code for IPs with AXI4 interfaces. This generated code of the IP having an AXI4 interface can be used to create peripherals in Vivado, a popular tool for designing and implementing digital circuits. The generated AXI4 IP having port names compatible with the standard port names of AXI4 peripherals in Vivado can make it easy to automate the connection process in the Vivado block design, which automatically connects the IP with the MicroBlaze/Zynq processor through AXI-Interconnect.

\section{Example Case Studies}

This section offers a diverse collection of case studies comparing Python DSL implementations with hand-crafted RTL and HLS-based implementations. 

\subsection{Post Quantum Cryptography}

This case study illustrates the usage of Python DSL to implement primitives for Post Quantum Cryptography (PQC). PQC refers to new cryptography schemes that are resistant to attacks from quantum computers. CRYSTALS-Kyber \cite{kyber} is a Lattice-based PQC scheme, which depends on the hardness of the Module-Learning-with-Errors (M-LWE) problem. Kyber is one of the winners in the NIST PQC Standardization competition and is being integrated into libraries and systems by the industry.\\

\noindent
The mathematical objects in CRYSTALS-Kyber are polynomials over $R_q = \mathcal{Z}_q[x]/(x^n + 1)$, where $q$ is 3329, $n$ is 256 and $k$ (which is the number of polynomials used) is 2 or 3 or 4. Hence, computations in Kyber involve polynomial matrix-matrix products and matrix-vector products. It also requires pseudo-random number generators.\\ 
Key building blocks for Kyber computations are Number Theoretic Transform (NTT) and Keccak-f[1600] core. Because polynomial products are compute-intensive (time complexity is $O(n^2)$), but if performed in the NTT domain, they are way more efficient (time complexity reduces to $O(n \log{n})$). And Keccak core is used in pseudo-random number generation.\\

\noindent
Various methods exist for implementing an NTT unit \cite{ntt}. For Kyber, 256-point NTT is to be computed by using two independent 128-point NTTs. Here, FIFOs are used to receive inputs and send outputs. The data is stored in registers (instead of BRAMs) for parallel memory access. Pipelining is done by running multiple ForLoopSections under a ParallelSection as shown. A single LeafSection is present under each ForLoopSection (describing a single pipeline stage).

\begin{footnotesize}
\begin{verbatim}
with ParallelSections ("PS_1):
    with ForLoopSection ("FLS_1, "i", 0, N):
        with LeafSection ("LS_1):
            ...stage 1 computation...
    with ForLoopSection ("FLS_2, "j", 1, N+1):
        with LeafSection ("LS_2):
            ...stage 2 computation...
    with ForLoopSection ("FLS_3, "k", 2, N+2):
        with LeafSection ("LS_3):
            ...stage 3 computation...
\end{verbatim}
\end{footnotesize}

\noindent
For Vivado HLS implementation, pipeline pragma is used. Inverse NTT is described similarly, followed by an extra pipeline in the end for multiplying with $n^{-1}$.\\

\noindent
The Keccak-f[1600] function (also called block transformation) involves 5 steps ($\theta$, $\rho$, $\pi$, $\chi$ and $\iota$) \cite{sha3}.
The computation takes 24 iterations (described using ForLoopSection) and each iteration takes 9 clock cycles (i.e. 9 LeafSections). Static elaboration of Python for-loop is used under each LeafSection to perform operations in parallel. For Vivado HLS implementation, unroll pragma is used.\\

\noindent
Table \ref{nttcc} compares Vivado HLS and Python DSL implementations of the mentioned hardware units. All the units generated by both tools can run at a clock frequency of $125~MHz$ on PYNQ-Z2 FPGA.

\begin{table}[!ht]
 \centering
\renewcommand{\arraystretch}{1.25} 
\begin{tabular}{l|l|l|l}
     & Vivado HLS & \begin{tabular}[c]{@{}l@{}}Python DSL\\ (no pipeline)\end{tabular} & \begin{tabular}[c]{@{}l@{}}Python DSL\\ (pipeline)\end{tabular} \\ \hline
NTT-256         & 3141       & 3136    & 454         \\
INTT-256        & 3272       & 3904    & 587         \\
Keccak-f[1600]  & 673   & 216      & -
\end{tabular}
\vspace{0.15cm}
\caption{PQC blocks using Vivado HLS and Python DSL (clock cycles)}
\label{nttcc}
\end{table}

\subsection{Matrix Multiplication}

The next example presents an integer matrix multiplier design utilizing \textit{SimpleFIFOs} and \textit{SimpleBRAMs}, along with a single pipelined multiply-add (\textit{MAC}) unit. It serves as a test-bed to compare the performance of Python DSL-generated Verilog code against manually written Verilog code for the same matrix multiplication algorithm across square matrices of various dimensions. Both implementations maintain identical interfaces and follow the same computational steps. The comparison aims to understand the trade-offs between these coding methods using various metrics. The multiplication involves two matrices, \( A_{N \times Q} \) and \( B_{Q \times M} \), resulting in a product matrix \( C_{N \times M} \). The study focuses on square matrices where \( N = Q = M \).

\subsubsection*{Initialization of SimpleBRAMs} 

Matrices $A_{N \times Q}$ and $B_{Q \times M}$ are fed to \textit{SimpleBRAMs} through \textit{InputFIFOs} and SimpleBRAM C is initialized with zeros in parallel. 


\noindent
Three \textit{ForLoopSection} (R\_A, R\_B and R\_C) are ran in parallel using \textit{withParallelSections} (par\_A\_B\_C) construct to effectively use potential parallelism for initializing \textit{SimpleBRAMs}

\begin{footnotesize}
\begin{verbatim}
 1    with ParallelSections ("par_A_B_C"):
 2     with ForLoopSection ("R_A", "p", 0, N * Q):
 3      with LeafSection ("recv_A"):
 4       A.writeData (p, fA.read ())
 5     with ForLoopSection ("R_B", "q", 0, Q * M):
 6      with LeafSection ("recv_B"):
 7       B.writeData (q, fB.read ())
 8     with ForLoopSection ("R_C", "r", 0, N * M):
 9      with LeafSection ("Initialize_C"):
 10      C.writeData (r, Const (32, 0))
\end{verbatim}
\end{footnotesize}

\subsubsection*{Matrix Multiplication} 

After initialization of \textit{SimpleBRAMs}, by accessing specific addresses, the data from \textit{SimpleBRAMs} will be enqueued into \textit{MAC} and the result of \textit{MAC} will be stored in \textit{SimpleBRAM C}. The following algorithm will be followed.

\begin{algorithm}[H]
\caption{Matrix Multiplication}
\begin{small} 
\begin{algorithmic}[1]
\Require Initialized $A$, $B$, and $C$ SimpleBRAMs
\For{$k = 0$ to $Q-1$}
  \For{$i = 0$ to $N-1$}
    \For{$j = 0$ to $M-1$}
      \State $C[i*M + j] \gets A[i*Q + k] \cdot B[k*M + j] + C[i*M + j]$
    \EndFor
  \EndFor
\EndFor
\end{algorithmic}
\end{small} 
\end{algorithm}

\begin{figure}
    \centering
    \includegraphics[height=3.5cm]{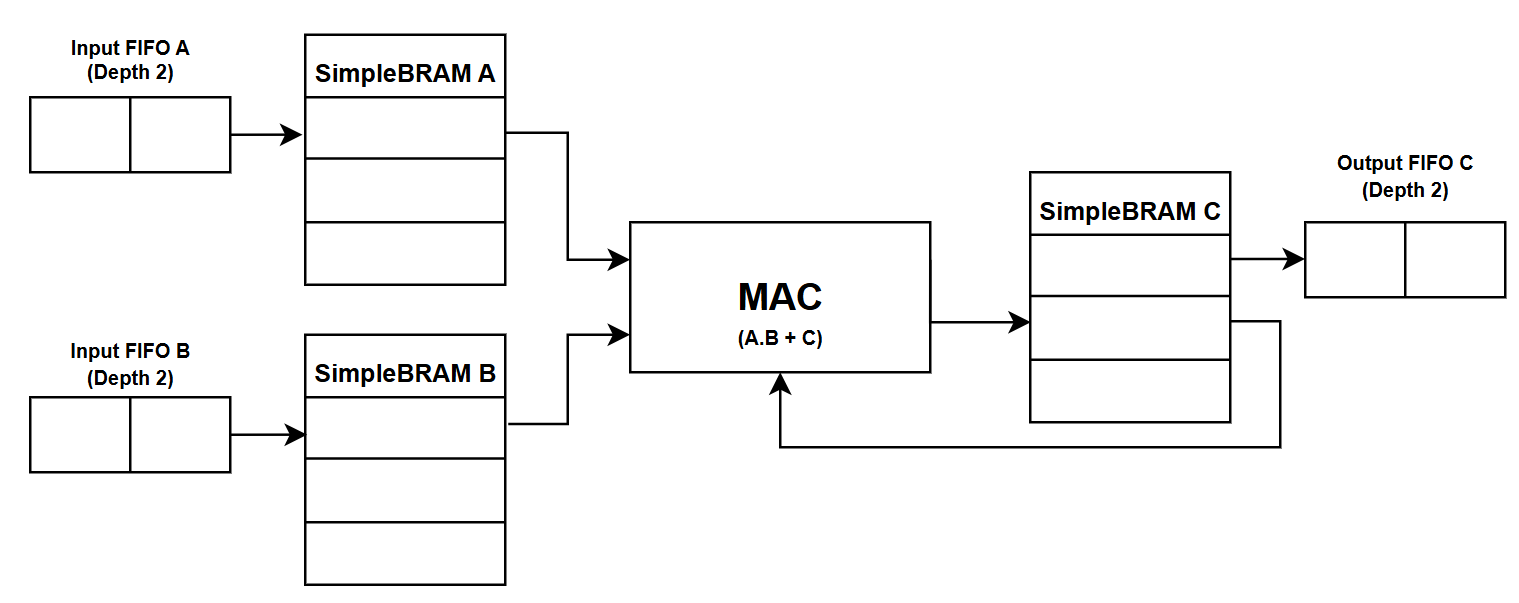}
    \caption{Matrix Multiplication Hardware}
    \label{fig:mmh}
\end{figure}

\noindent
Refer to Figure \ref{fig:mmh} for the architecture of the matrix multiplier.\\

\subsubsection*{Code Efficiency}
Comparing the number of lines in the Python DSL-generated code versus the manually written code, the DSL-generated Verilog code is approximately \( 43.94\% \) greater in size than the manually written Verilog code.

\begin{table}[!ht]
    \centering
    \renewcommand{\arraystretch}{1.25} 
    \begin{tabular}{c|c}
        Code & Lines of Code  \\ \hline
        DSL generated Verilog code & 416  \\
        Manually written Verilog code & 289  \\
    \end{tabular}
    \vspace{0.15cm}
    \caption{Code Efficiency}
    \label{CodeEfficiency}
\end{table}

\subsubsection*{Performance}

\begin{figure}[!ht]
    \centering
    \includegraphics[scale=0.32]{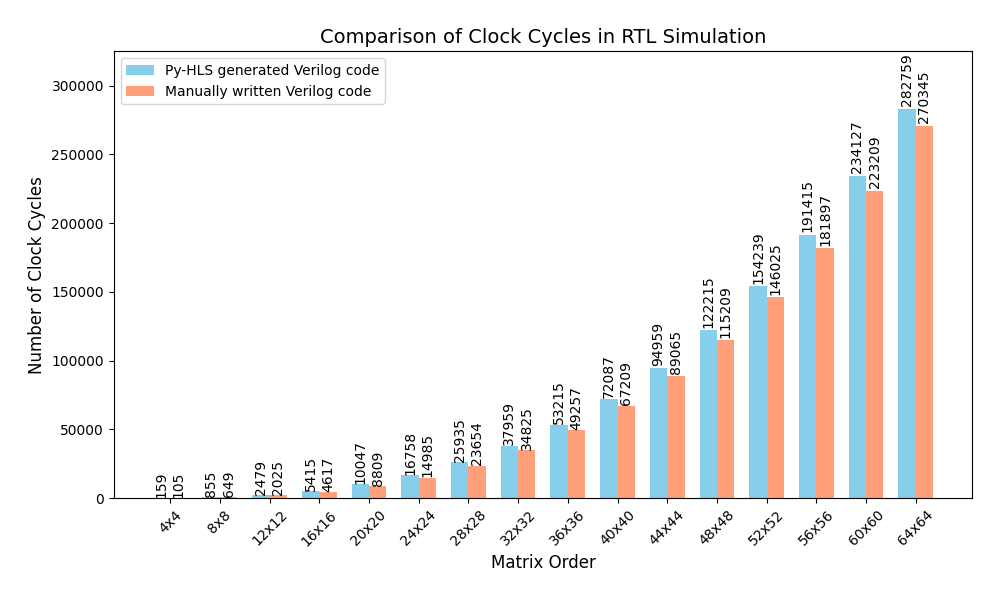}
    \caption{Clock Cycles taken for RTL Simulation}
    \label{fig:mmul_Performance}
\end{figure}

To measure the execution time for RTL Simulation, both codes were tested using the same Verilog testbench, and the number of clock cycles between the assertion of the start signal and the assertion of the done signal were compared for both codes for matrix multiplication of square matrices of various dimensions. The number of clock cycles is plotted against the matrix dimensions as shown in Figure \ref{fig:mmul_Performance}.\\

\subsubsection*{Resource Utilization}

\begin{figure}[!ht]
    \centering
    \includegraphics[scale=0.23]{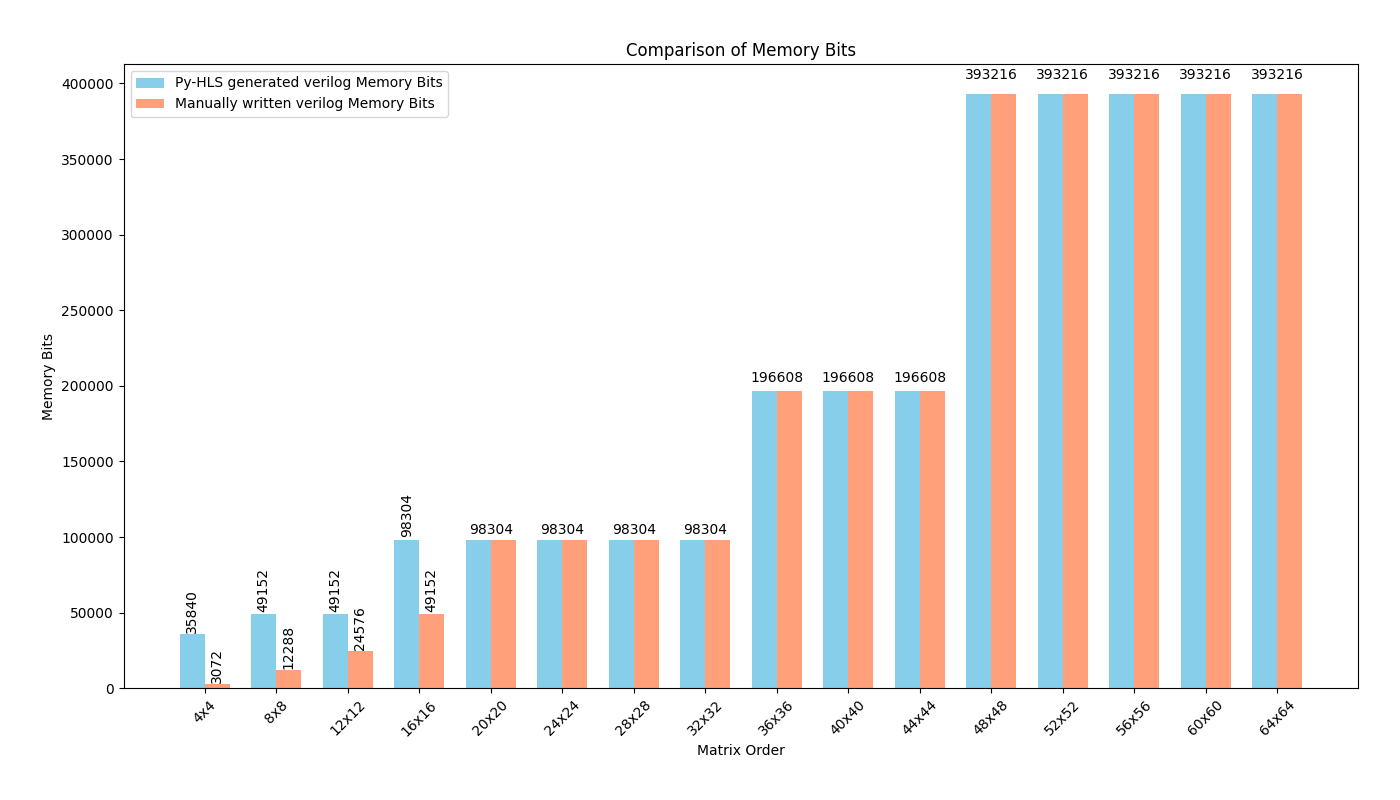}
    \caption{Comparison of Memory Bits}
    \label{fig:Memory Bits Comparison}
\end{figure}

\begin{figure}[!ht]
    \centering
    \includegraphics[scale=0.23]{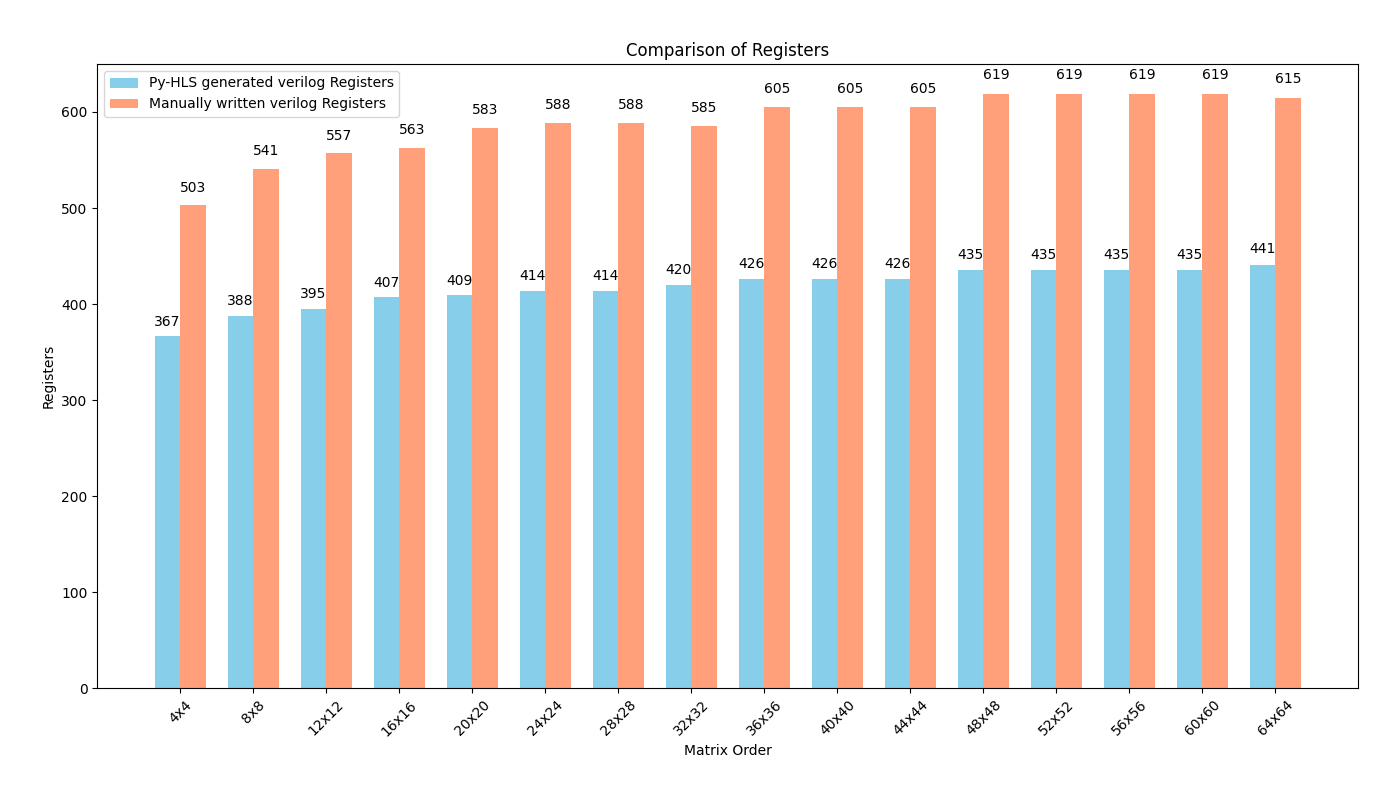}
    \caption{Comparison of Number of Registers}
    \label{fig:Number of Registers Comparison}
\end{figure}

\begin{figure}[!ht]
    \centering
    \includegraphics[scale=0.23]{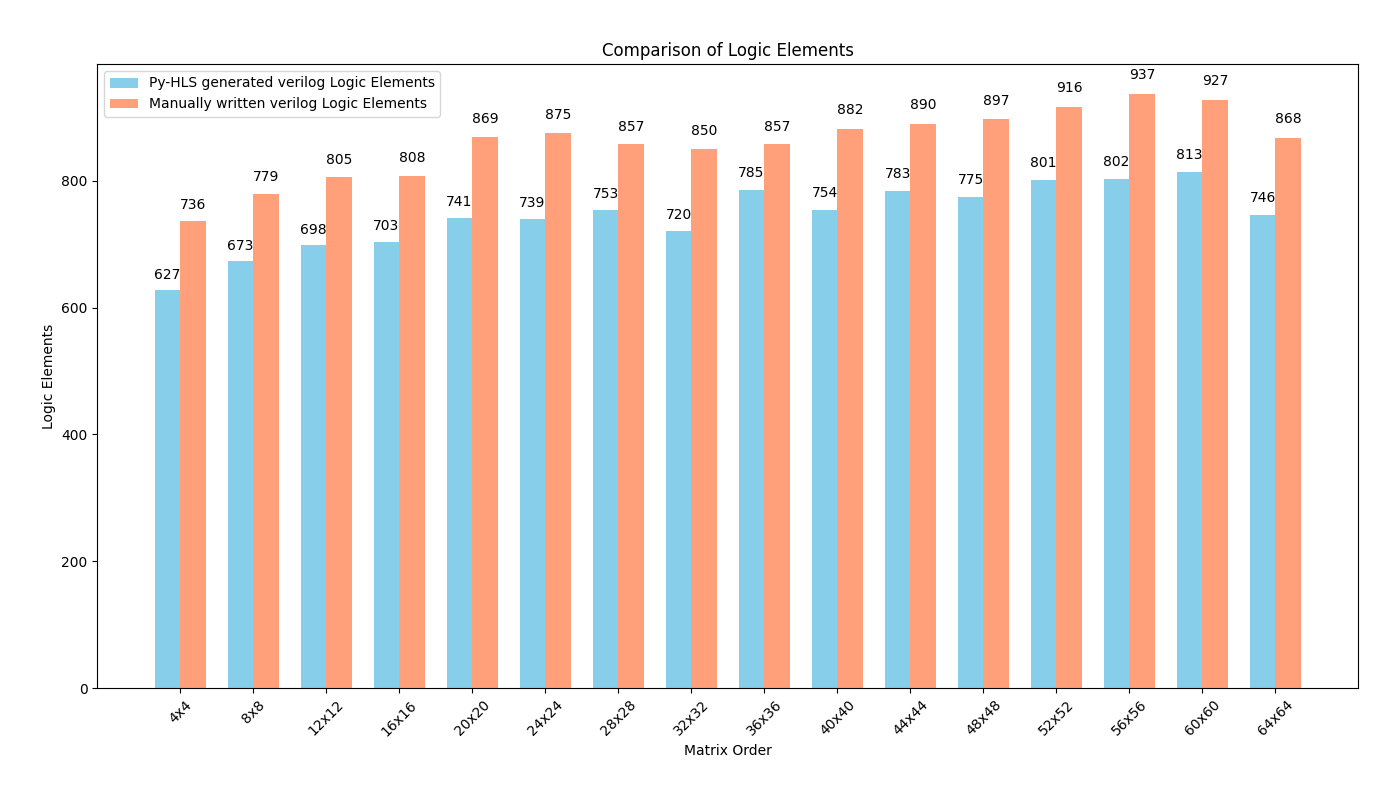}
    \caption{Comparison of Number of Logic Elements}
    \label{fig:Number of Logic Elements Comparison}
\end{figure}

The hardware resource utilization for matrix multiplication on the EP4CE22F17C6 device (De0-Nano FPGA Board) was analyzed and compared for Python DSL-generated and manually written Verilog codes. Compilation reports from Quartus Prime Lite 18.1 were used for this analysis. The utilization of resources such as logic elements, flip-flops, and BRAMs was plotted for matrix multiplication of square matrices of various dimensions. Figures \ref{fig:Memory Bits Comparison}, \ref{fig:Number of Registers Comparison}, and \ref{fig:Number of Logic Elements Comparison} depict the comparisons.


\subsection{Stereo-vision}

Stereo-vision is an imaging technique used to obtain 3D measurements of an arbitrary scene, using 2D images of the same scene captured from different viewpoints using a stereo camera. The algorithm uses the principle of triangulation method. Real-time stereo-vision processing of the scene involves calibration of the captured images, finding correspondence (disparity) between captured images, calculating depth using disparity values, and imaging configuration geometric parameters.\\
Semi-global mapping \cite{sgm, SingleSS} is used in stereo vision for dense disparity map computation which helps to obtain improved accuracy in estimating disparities (depth information) between corresponding points in stereo images over local block-matching techniques. SGM uses a dynamic programming approach to optimize the energy cost function of a pixel. This work presents a variant of SGM called MGM-4 (More Global Matching) wherein neighbouring pixel disparity variations along four directions (top, top-left, top-right, left) are considered. For stereo-image inputs (Image height – $H$, Image Width – $W$) and search range - $D$, Table \ref{sgmopt} illustrates how SGM is optimally implemented.

\begin{table}[ht!]
\renewcommand{\arraystretch}{1.25} 
\centering
\begin{tabular}{l|l|l}
S.No & Steps used in SGM & 
\begin{tabular}[c]{@{}l@{}} Execution count \end{tabular} \\ \hline
1 & Input pixel reading  & $H \times W$  \\ \hline
2 & Matching cost computation & $H \times W \times D$ \\ \hline
3 & \begin{tabular}[c]{@{}l@{}} Path cost computation\\  (along different paths)\end{tabular} 
& $4 \times H \times W \times D$  \\ \hline
4 & Average sum cost computation                   & $H \times W \times D$ \\ \hline
5 & Disparity assignment & $H \times W$  
\end{tabular}
\caption{SGM Optimization}
\label{sgmopt}
\end{table}

\noindent
Step 3 is executed a maximum number of times for the disparity computation of one image pair. As the input image size scales up, computational complexity increases. Hence, cost optimisation along 4 paths can be computed in parallel, reducing time complexity from ($4 \times H \times W \times D$) to $H \times W \times D$. All the steps are inter-dependent on their previous step except for step 2 and step 3; hence, step 2 and step 3 can also be parallelized.\\

\noindent
Python DSL implementation of SGM is elaborated below.
\begin{itemize}
\item Two input image pixel values are sent using FIFOs and are stored in registers for internal computations.
\vspace{0.1cm}
\item Internal buffers and arrays are initialized in parallel.
\begin{footnotesize}
\begin{verbatim}
with ParallelSections ("P1"):
    with ForLoopSections ("inp","v0",0,size):
        with LeafSection ("LS_1):
            ...read pixels from input FIFOs... 
    with ForLoopSections ("bf,"v1",0,size):
        with LeafSection ("LS_2):
            ...initialize buffer and array...   
\end{verbatim}
\end{footnotesize}

\item Correspondence values are computed pixel by pixel using SGM, by calculating disparity volume and disparity decision. 
\begin{footnotesize}
\begin{verbatim}
with ForLoopSections ("r",v1,0,height):
    with ForLoopSections ("c",v2,0,width):
        ...update buffers and arrays...
        with ForLoopSections ("d",v3,0,range):
            ...matching cost calculation...
            with LeafSection ("path_cost_cal"):
                for var4 in range(n_dir)
                    ..compute path_cost...
            ...compute sum_cost...
            ...compute avg(MGM-4)...
            ..disparity decision...
    ...assign the final disparity...
\end{verbatim}
\end{footnotesize}

\item Output disparity is written into FIFO and is read from it for depth calculation as a final step.
\end{itemize}

\noindent
Table \ref{sgmresults} compares Python DSL implementation against Vivado HLS and PandA-bambu HLS \cite{bambu} implementations of SGM (input images of size $8 \times 8$, filter size $3\times3$ and search range $5$). All units are synthesized on PYNQ-Z2 FPGA and can run at a clock frequency of $100~MHz$.

\begin{table}[h!]
\centering
\renewcommand{\arraystretch}{1.25} 
\begin{tabular}{l|l|l}
Tool  & Design Model & Clock Cycles\\ \hline
\multicolumn{1}{c|}{\multirow{2}{*}{Vivado HLS}} & Without pragmas & 11833  \\ \cline{2-3} 
\multicolumn{1}{c|}{}                            & \begin{tabular}[c]{@{}l@{}}With parallel\\ cost computations\end{tabular}  & 2021  \\ \hline
\multirow{2}{*}{bambu HLS} & \begin{tabular}[c]{@{}l@{}}Without any\\ optimizations\end{tabular} & 9080   \\ \cline{2-3} 
& \begin{tabular}[c]{@{}l@{}}With level 3\\ compiler optimization\end{tabular}    & 1994 \\ \hline
\multirow{2}{*}{Python DSL}
& \begin{tabular}[c]{@{}l@{}}With parallel\\ cost computation\end{tabular}    & 1247  \\ 
\end{tabular}
\vspace{0.15cm}
\caption{SGM using Vivado HLS, bambu HLS and Python DSL}
\label{sgmresults}
\end{table}

\subsection{Fast Fourier Transform}

\begin{figure}[!ht]
        \centering
        \includegraphics[scale=0.205]{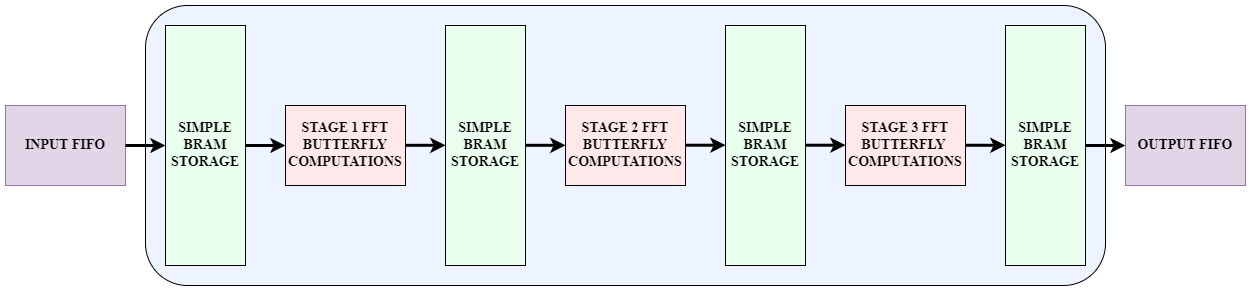}
        \caption{FFT Hardware Architecture}
        \label{8-point FFT Arch}
\end {figure}

FFT is arguably the most widely used algorithm in signal processing. It is used to obtain the frequency-domain spectrum of a time-domain signal \cite{fft}. In this case study, 256-point FFT is designed using the Python DSL with FIFO interface. Inputs are taken through input FIFOs and stored in BRAMs. This FFT implementation uses the Cooley-Tuckey Algorithm. 256-point FFT requires 8 stages. After each computation stage, outputs are stored in BRAMs and the next stage takes inputs from these BRAMs. The final outputs are taken out using output FIFO. For illustration, Figure \ref{8-point FFT Arch} shows the hardware architecture of 8-point FFT implemented using Python DSL. The same architecture has been extended to compute 256-point FFT.

\noindent
The Python DSL implementation takes 12606 clock cycles to perform 256-point FFT. The same design using Vivado HLS without exploiting parallelism and pipelining takes 12897 clock cycles for the computation. Both are synthesized on PYNQ-Z2 FPGA with a target clock frequency of $100~MHz$.

\subsection{Discrete Wavelet Transform}
DWT is an important algorithm used in signal and image processing applications. Image compression is one of the prominent applications that use DWT. DWT can be computed using various kernel functions the simplest one being the Haar wavelet \cite{haardwt}. This wavelet involves averaging and different operations to compress the image.

\begin{figure}[!ht]
        \centering
        \includegraphics[scale=0.15]{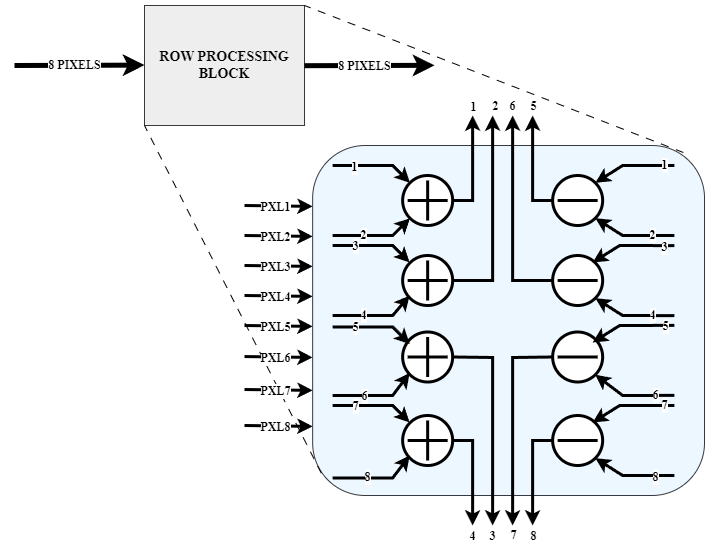}
        \caption{DWT Row Processing Module}
        \label{DWT Row Arch}
\end {figure}

\begin{figure}[!ht]
        \centering
        \includegraphics[scale=0.15]{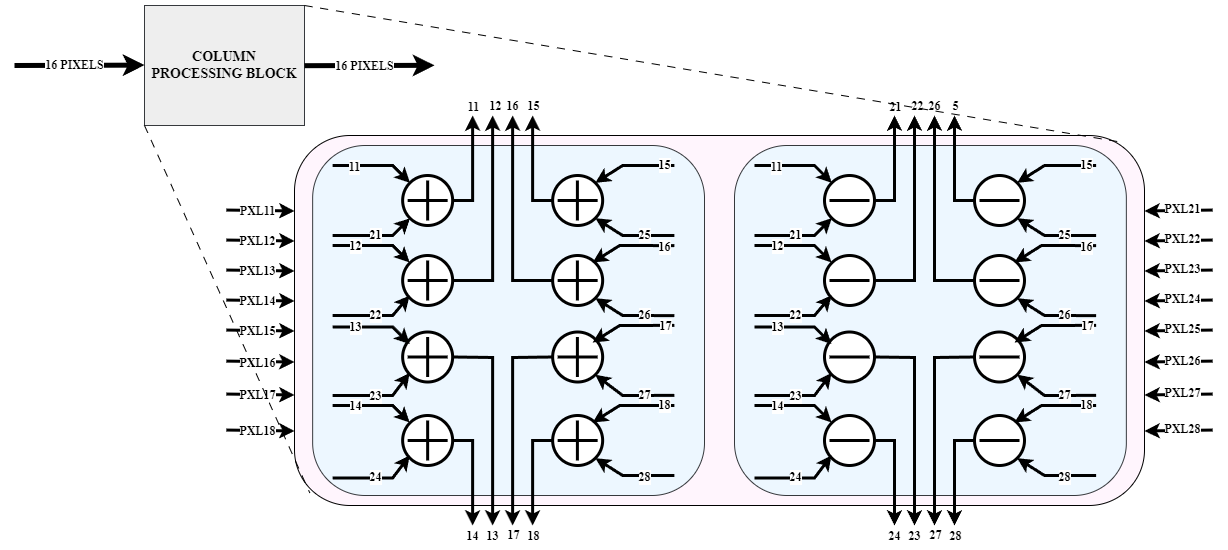}
        \caption{DWT Column Processing Module}
        \label{DWT Col Arch}
\end {figure}

\noindent 
The architecture of the image compression consists of a row-processing module and a column-processing module. These modules perform averaging and differencing operations row-wise and column-wise respectively on the input pixels. The image is divided into processing blocks of 16 pixels each which are enqueued into input FIFOs and after the processing dequeued from output FIFOs. The intermediate pixel values are stored in BRAMs after the row-processing module. Figures \ref{DWT Row Arch} and \ref{DWT Col Arch} show the architectures of the row and column-processing modules used in the Python DSL implementation of DWT.

\noindent
The Python DSL implementation takes approximately 10560 clock cycles to compress an image of $32\times32$ pixels.

\subsection{Digital Correlator}
Digital correlator is widely used in signal processing applications such as detecting characteristics of an input signal with respect to reference signals. One of the applications of this is to detect the signal amplitude at particular frequencies from an input signal comprising different frequencies, by calculating the maximum value of correlation with the reference signal. \\
In Python DSL, the above correlation application is modelled using the linear buffering algorithm. The flipped version of the reference signal is stored in BRAM using an input FIFO. Based on its frequency, 37 samples are stored based on the sampling frequency covering one period of the reference signal of uint type for 32-bit fixed point representation. In the linear buffering algorithm, input data is received sequentially using an input FIFO. The linear buffer, which has a size of 512, is shifted, storing the latest sample in the initial position. Accumulation is then performed with the reference signal. The correlator hardware takes 3586 clock cycles. For hardware running at $100~MHz$, the design can run on input signals with a maximum of $27.8~kHz$ sampling frequency.\\

\begin{figure}[h!]
\centering {\includegraphics[scale=0.23]{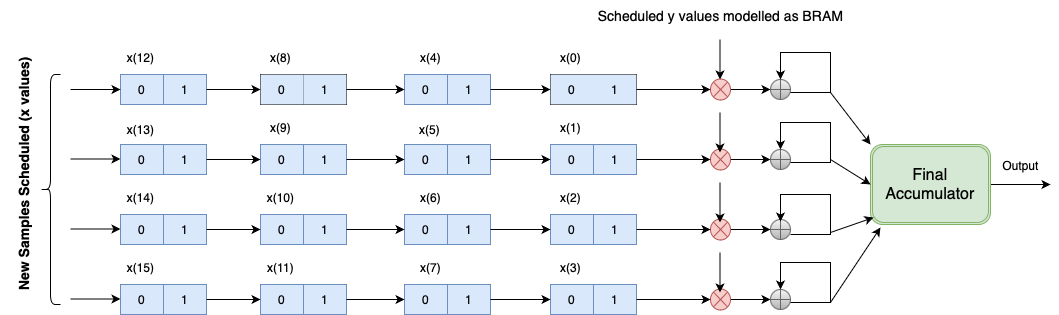}}
    \caption{Folding correlator architecture}
    \label{Folding correlator architecture}
\end{figure}

\noindent
The Python DSL tool can be used to parallelize such designs by implementing folded architectures. Figure \ref{Folding correlator architecture} shows a folded-by-4 architecture for a sample size of 16 as a test example which can be scaled to large sizes. The architecture uses a chain of FIFOs (BRAMs as FIFOs) to schedule the inputs and reference signals.

\subsection{Butterfly Mating Optimization Algorithm}

\begin{figure}[h!]
\centering {\includegraphics[scale=0.23]{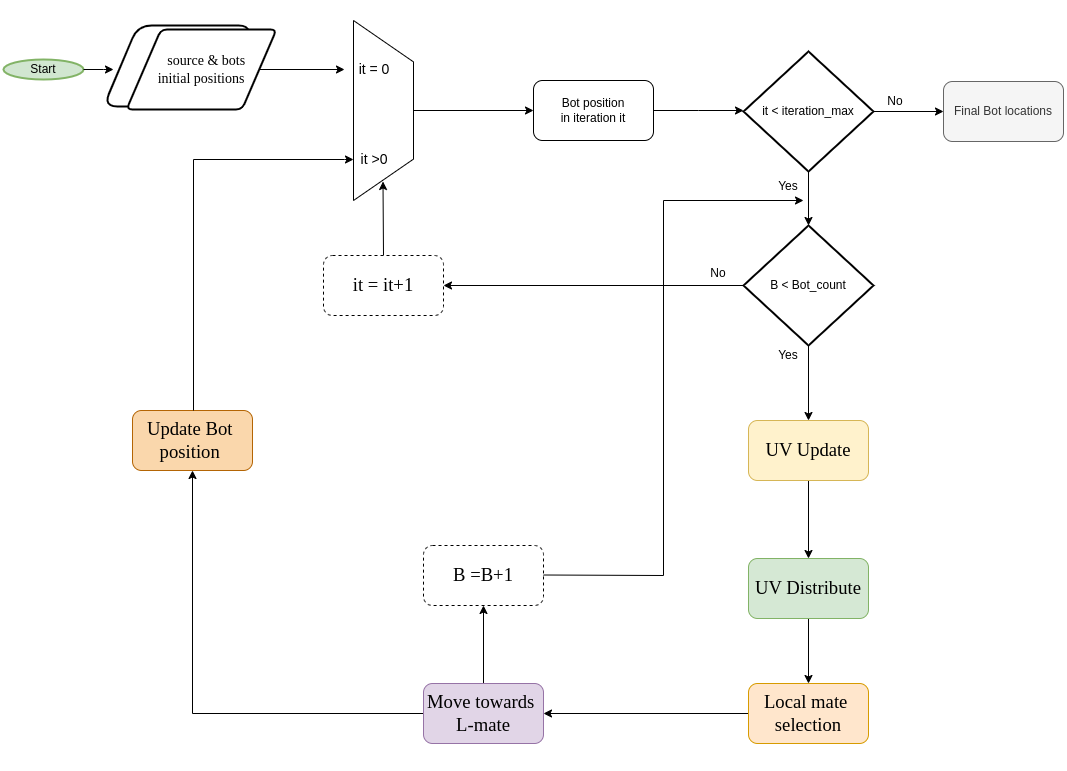}}
    \caption{Flowchart of BMO Algorithm}
    \label{Algorithm flowchart of BMO}
\end{figure}

Inspired by mating behavior in birds, the BMO algorithm \cite{bmo2} tackles complex optimization problems with multiple possible solutions. The algorithm achieves this by introducing ``Bot" entities that explore the search space in place of the traditional ``natural butterfly" concept.
The algorithm starts with a random initialization of Bots ($x,y$ coordinates) in the search space aimed at reaching the target position. Each Bot has its self UV value updated from the UV updating phase of the algorithm based on the positions of the Bots. After UV updating, each Bot distributes its updated UV to the remaining Bots in the UV distribution phase such that the nearest Bot gets more than the farthest one. Once the Bot receives multiple UVs from the distribution phase, it searches for the maximum UV value distributed Bot (local mate). The Bot then moves towards the local mate by updating its position in the movement phase based on the bot's step size. Figure \ref{Algorithm flowchart of BMO} illustrates the core aspects of the BMO algorithm and it aims to provide a comprehensive overview of certain optimizations. The following pseudo-code (Algorithm \ref{alg:cap}) outlines the key steps.

\begin{algorithm}
\begin{algorithmic}
\Require Number of Bots (Bot$_-$count),\\
            Pre-initialized Bots \& Source Locations 
\Ensure Set all bots UV$_i$ = 0 
\State Number of iterations $\gets$ iteration$_-$count
\State $it \gets 0$
\State $B \gets 1$
\While{it $\leq$ iteration$_-$count}
    \While{B $\leq$ Bot$_-$count}
        \State Update UV 
        \State Distribute UV
        \State L-mate selection
        \State Position Update
        \State $B \gets B+1$
    \EndWhile
\State $it \gets it+1$
\EndWhile
\end{algorithmic}
\caption{BMO algorithm}\label{alg:cap}
\end{algorithm}

\noindent
The computed positions of 4-Bots for 40 iterations achieve an accuracy of 98.2\% (Refer Table \ref{Clock Cycle Count for Different Number of Bots}). Increasing the iteration count above 40 will not improve the accuracy much for a particular step-size. 
Table \ref{Comparision between Implementation of BMO algorithm using Vivado HLS and PyHLS} shows the Synthesis Reports and Cycle Count of implemented BMO algorithm for 4-Bots using Python DSL and Vivado HLS (without pipeline pragma) 
on targeting ZCU104 board.


\begin{table}[!ht]
    \centering
    \renewcommand{\arraystretch}{1.25} 
    \begin{tabular}{c|c|c|c}
        Number of Bots & Iterations & Clock Cycles & Error \\ \hline
        2 & 20 & 852  & 1.2\%\\
        4 & 20 & 852  & 12.2\% \\
        4 & 40 & 1692 & 1.8\% \\
    \end{tabular}
    \vspace{0.15cm}
    \caption{Clock Cycle Count for Different Number of Bots}
    \label{Clock Cycle Count for Different Number of Bots}
\end{table}

\newpage
\begin{table}[!hb]
    \centering
    \renewcommand{\arraystretch}{1.25} 
        \begin{tabular}{c|c|c}
        Parameter & Python DSL & Vivado HLS  \\ \hline
        Clock Cycles & 2438  & 321553 \\
        Frequency of Operation & $100~MHz$  & $100~MHz$ \\
        Co-Simulation Latency & $24~\mu s$  & $3.21~ms$ \\
        FF & 1\%  & 2\% \\
        LUT & 2\%  & 4\% \\
        DSP & 0\%  & 2\% \\
        BRAM & 0\% & 1\% \\
        \end{tabular}
    \vspace{0.15cm}
    \caption{BMO algorithm using Vivado HLS and Python DSL}
    \label{Comparision between Implementation of BMO algorithm using Vivado HLS and PyHLS}
\end{table}

\section{Future Work}

Some interesting extensions and potential additional features are outlined in this section. These can be developed in the existing DSL framework itself.

\subsection{Detecting conflicting updates}

Static code analysis frameworks such as Frama-C can estimate a variable's possible set of values when execution reaches a given line of source code.
This tool can be used to detect if two LeafSection blocks that are updating the same state element can be active at the same clock cycle. Further, one may think of automatically breaking a LeafSection into multiple LeafSections.

\subsection{Breakpoints in Hardware}

Debuggers such as {\tt gdb} are immensely useful to software developers. A debugger allows tracing execution of a program, often without even recompiling the code. As HLS tools allow users to express the model in a higher-level language such as C/C++/Python, it will be interesting if the user is also allowed to set a breakpoint at a line of input source code, and when the update operation specified by that line is active in hardware, the hardware can be paused, allowing users to inspect the values of all other state elements. This feature would be useful when the entire HDL model cannot be simulated to debug interfacing problems with I/O, sensors, etc.

\subsection{Dynamic memory allocation}
HLS tools are typically used to design large parts of an entire application (as opposed to RTL design -- where module-by-module low-level design
is done). These applications are run on a reconfigurable SoC, e.g. ARM + FPGA (such as Xilinx Zynq, Intel Stratix), or with Microblaze (Xilinx) / Nios (Intel) CPUs implemented in FPGA itself. Interconnects such as AXI/Avalon and compatible DDR controller modules make it easy for hardware IPs to access DDR DRAM. Applications may need dynamic memory allocation. It is better to leave the DDR memory management to the CPU. It typically involves maintaining boundary tags, various flags, free lists for various-sized buffers, etc. Hardware modules could be made to interrupt the CPU with the allocation/de-allocation request, and the CPU can manage memory that can be used by the hardware as it continues. This feature can be prototyped in this DSL easily.

\section{Conclusion}

In this paper, a Python DSL for generating Verilog model of synchronous digital circuits is introduced. Details on how the DSL has been constructed and how to use it have been provided. Various case studies have been made, illustrating how the Python DSL can be used for rapid prototyping of complicated hardware. Finally, ideas for extending the DSL have been provided. The source codes for all the case studies presented in this paper are on the GitHub repository: \url{https://github.com/HPC-Lab-IITB/Python-DSL}. The source code for the Python DSL will be released upon publishing of the paper.

\bibliographystyle{IEEEtran} 
\bibliography{main}

\end{document}